\documentclass{emulateapj}

\begin{document}

\slugcomment{Accepted by the Astronomical Journal, March 24, 2009}
\shorttitle{Satellites of Haumea}
\shortauthors{Ragozzine \& Brown}

\title{Orbits and Masses of the Satellites 
of the\\ Dwarf Planet Haumea = 2003 EL61}
\author{D. Ragozzine and M.~E. Brown}
\affil{Division of Geological and Planetary Sciences, California Institute
of Technology, Pasadena, CA 91125}

\email{darin@gps.caltech.edu}



\newcommand{\elso}{2003~EL61 }
\newcommand{\elsons}{2003~EL61}
\newcommand{\Nfive}{2005 CB79 }
\newcommand{\Kfive}{2005 UQ513 }
\newcommand{\x}{\times}
\newcommand{\e}[1]{10^{#1}}
\newcommand{\be}{\begin{equation}}
\newcommand{\ee}{\end{equation}}
\newcommand{\mpers}{m s$^{-1}$}
\newcommand{\mpersws}{m s$^{-1}$ }

\newcommand{\bfam}{B07 }
\newcommand{\bfamnat}{2007Nature..446..296}
\newcommand{\bfamns}{B07}

\newcommand{\fyninet}{1}
\newcommand{\barkumet}{2}
\newcommand{\bfamt}{3}
\newcommand{\mbosst}{4}
\newcommand{\nollpostert}{5}
\newcommand{\teglert}{6}


\begin{abstract}

Using precise relative astrometry from the Hubble Space Telescope and the W. M. Keck Telescope, we have determined the orbits and masses of the two 
dynamically interacting satellites of the dwarf planet (136108) Haumea, formerly \elsons. The orbital parameters of Hi'iaka, the outer, brighter 
satellite, match well the previously derived orbit. On timescales longer than a few weeks, no Keplerian orbit is 
sufficient to describe the motion of the inner, fainter satellite Namaka. Using a fully-interacting three point-mass model, we have recovered the orbital parameters 
of both orbits and the mass of Haumea and Hi'iaka; Namaka's mass is marginally detected. The data are not sufficient to uniquely determine the gravitational 
quadrupole of the non-spherical primary (described by $J_2$). The nearly co-planar nature of the satellites, as well as an inferred density similar to water 
ice, strengthen
the hypothesis that Haumea experienced a giant collision billions of years ago. The excited eccentricities and mutual inclination point to an 
intriguing tidal history of significant semi-major axis evolution through satellite mean-motion resonances. The orbital solution indicates that Namaka 
and Haumea are currently undergoing mutual events and that the mutual event season will last for the next several years.

\end{abstract}

\section{Introduction} 

The dwarf planet (136108) Haumea, formerly \elsons, and about 3/4 of other large Kuiper belt objects (KBOs) have at least one small close-in satellite 
\citep{2006Natur.439..943W,2006ApJ...639L..43B,2007IAUC.8812....1B}. All of these larger KBOs are part of the excited Kuiper belt, where the detectable 
binary fraction among smaller KBOs is much lower, only a few percent \citep{2006AJ....131.1142S}. In contrast, the cold classical population (inclinations $\lesssim 5^{\circ}$) has 
no large KBOs \citep{2001AJ....121.1730L,2008ssbn.book..335B}, but prevalent widely separated binaries with nearly equal masses \citep{2008Icar..194..758N}. The 
differences between the types and frequency of Kuiper belt binaries may point to different binary formation mechanisms. Small satellites of large KBOs 
appear to be formed by collision, as proposed for the Pluto system \citep{2005Sci...307..546C,2006Natur.439..946S}, Eris and Dysnomia 
(\citealp{2007Sci...316.1585B}, but see \citealp{2008Icar..194..847G}), and Haumea \citep{2006ApJ...640L..87B,2007Nature..446..296,fraserbrowninprep}, 
but smaller KBO binaries have more angular momentum than can be generated in typical impacts and are apparently formed by some other mechanism 
\citep[e.g.,][]{2002Icar..160..212W,2002Natur.420..643G,2004Natur.427..518F,2005MNRAS.360..401A,2008DPS....40.3802N}. Both mechanisms of binary 
formation require higher number densities than present in the current Kuiper belt, as modeled explicitly for the Haumea collision by 
\citet{2008AJ....136.1079L}.

The collisional origin of Haumea's two satellites --- the outer, brighter satellite Hi'iaka (S1) and the inner, fainter satellite Namaka (S2) ---
is inferred from several related observations. Haumea has a moderate-amplitude light-curve and the 
shortest rotation period (3.9155 hours) among known objects of its size \citep{2006ApJ...639.1238R}. The rapid rotation requires a large spin angular 
momentum, as imparted by a large oblique impact. Using the mass of Haumea derived by the orbit of Hi'iaka \citep[][hereafter 
B05]{2005ApJ...632L..45B}, assuming Haumea's rotation axis is nearly perpendicular to the line-of-sight (like the satellites' orbits), and assuming the 
shape is that of a Jacobi ellipsoid (a homogeneous fluid), the photometric light curve can be used to determine the size, shape, albedo, and density of 
Haumea (\citealp{2006ApJ...639.1238R}; \citealp{2007AJ....133.1393L}, but see \citealp{2007Icar..187..500H}). 
It is estimated that Haumea is a tri-axial ellipsoid with approximate semi-axes of 500 x 750 x 
1000 km with a high albedo (0.73) and density (2.6 g/cm$^3$), as determined by \citet{2006ApJ...639.1238R}. This size and albedo are consistent with 
Spitzer radiometry \citep{2008ssbn.book..161S}. The inferred density is near that of rock and higher than all known KBOs 
implying an atypically small ice fraction.

Haumea is also the progenitor of the only known collisional family in the Kuiper belt \citep{2007Nature..446..296}. It seems that the collision that 
imparted the spin angular momentum also fragmented and removed the icy mantle of the proto-Haumea (thus increasing its density) and ejected these 
fragments into their own heliocentric orbits. The Haumea family members are uniquely identified by deep water ice spectra and optically neutral color 
\citep{2007Nature..446..296}, flat phase curves \citep{2008AJ....136.1502R}, and tight dynamical clustering \citep{2007AJ....134.2160R}. The dynamical 
clustering is so significant that \citet{2007AJ....134.2160R} were able to correctly predict that 2003 UZ117 and 2005 CB79 would have deep water ice 
spectra characteristic of the Haumea family, as verified by \citet{2008ApJ...684L.107S}. The distribution of orbital elements matches the unique 
signature of a collisional family, when resonance diffusion \citep[e.g.,][]{2001Icar..150..104N} is taken into account. Using this resonance diffusion 
as a chronometer, \citet{2007AJ....134.2160R} find that the Haumea family-forming collision occurred at least 1 GYr ago and is probably primordial. 
This is consistent with the results of \citet{2008AJ....136.1079L}, who conclude that the Haumea collision is only probable between two scattered-disk 
objects in the early outer solar system when the number densities were much higher.

In this work, we have derived the orbits and masses of Haumea, Hi'iaka, and Namaka. In Section 2, 
we describe the observations used to determine precise relative astrometry. The orbit-fitting techniques and results are given in Section 3. Section 4 discusses the 
implications of the derived orbits on the past and present state of the system. We conclude the discussion of this interesting system in Section 5.

\section{Observations and Data Reduction}

Our data analysis uses observations from various cameras on the Hubble Space Telescope (HST) and the NIRC2 camera with Laser Guide Star Adaptive Objects at 
the W. M. Keck Observatory. These observations are processed in different ways; here we describe the general technique and below we discuss the individual 
observations. Even on our relatively faint targets ($V \approx 21,22$), these powerful telescopes can achieve 
relative astrometry with a precision 
of a few milliarcseconds. The Julian Date of observation, the relative astrometric distance on-the-sky, and the estimated astrometric errors are 
reported in Table \ref{obs}.


\begin{deluxetable*}{llllllllllll}
\tablecaption{Observed Astrometric Positions for the Haumea System \label{obs}}
\tabletypesize{\scriptsize}
\tablewidth{0pt}
\tablehead{
\colhead{Julian Date} & \colhead{Date} & \colhead{Telescope} & \colhead{Camera}  & \colhead{$\Delta x_H$} & \colhead{$\Delta y_H$} & \colhead{$\sigma_{\Delta x_H}$} & \colhead{$\sigma_{\Delta y_H}$}  & \colhead{$\Delta x_N$} & \colhead{$\Delta y_N$} & \colhead{$\sigma_{\Delta x_N}$} & \colhead{$\sigma_{\Delta y_N}$}  \\ 
& &  &  & arcsec & arcsec & arcsec & arcsec }
\startdata

  2453397.162 & 2005 Jan 26
 & Keck & NIRC2 &  0.03506  & -0.63055  &  0.01394  &  0.01394  & 
\nodata & \nodata & \nodata & \nodata \\
  2453431.009 & 2005 Mar  1
 & Keck & NIRC2 &  0.29390  & -1.00626  &  0.02291  &  0.02291  & 
 0.00992  &  0.52801  &  0.02986  &  0.02986  \\ 
  2453433.984 & 2005 Mar  4
 & Keck & NIRC2 &  0.33974  & -1.26530  &  0.01992  &  0.01992  & 
\nodata & \nodata & \nodata & \nodata \\
  2453518.816 & 2005 May 28
 & Keck & NIRC2 & -0.06226  &  0.60575  &  0.00996  &  0.00996  & 
\nodata & \nodata & \nodata & \nodata \\
  2453551.810 & 2005 Jun 30
 & Keck & NIRC2 & -0.19727  &  0.52106  &  0.00498  &  0.00996  & 
-0.03988  & -0.65739  &  0.03978  &  0.03978  \\ 
  2453746.525 & 2006 Jan 11
 & HST & ACS/HRC & -0.20637  &  0.30013  &  0.00256  &  0.00256  & 
 0.04134  & -0.18746  &  0.00267  &  0.00267  \\ 
  2453746.554 & 2006 Jan 11
 & HST & ACS/HRC & -0.20832  &  0.30582  &  0.00257  &  0.00257  & 
 0.03867  & -0.19174  &  0.00267  &  0.00267  \\ 
  2454138.287 & 2007 Feb  6
 & HST & WFPC2 & -0.21088  &  0.22019  &  0.00252  &  0.00197  & 
-0.02627  & -0.57004  &  0.00702  &  0.00351  \\ 
  2454138.304 & 2007 Feb  6
 & HST & WFPC2 & -0.21132  &  0.22145  &  0.00095  &  0.00204  & 
-0.03107  & -0.56624  &  0.00210  &  0.00782  \\ 
  2454138.351 & 2007 Feb  6
 & HST & WFPC2 & -0.21515  &  0.23185  &  0.00301  &  0.00206  & 
-0.03009  & -0.55811  &  0.00527  &  0.00564  \\ 
  2454138.368 & 2007 Feb  6
 & HST & WFPC2 & -0.21402  &  0.23314  &  0.00192  &  0.00230  & 
-0.03133  & -0.56000  &  0.00482  &  0.00663  \\ 
  2454138.418 & 2007 Feb  6
 & HST & WFPC2 & -0.21705  &  0.24202  &  0.00103  &  0.00282  & 
-0.03134  & -0.54559  &  0.00385  &  0.00376  \\ 
  2454138.435 & 2007 Feb  6
 & HST & WFPC2 & -0.21449  &  0.24450  &  0.00323  &  0.00254  & 
-0.02791  & -0.54794  &  0.00571  &  0.00524  \\ 
  2454138.484 & 2007 Feb  6
 & HST & WFPC2 & -0.21818  &  0.25301  &  0.00153  &  0.00224  & 
-0.02972  & -0.53385  &  0.00797  &  0.01330  \\ 
  2454138.501 & 2007 Feb  7
 & HST & WFPC2 & -0.21807  &  0.25639  &  0.00310  &  0.00291  & 
-0.03226  & -0.53727  &  0.00531  &  0.00400  \\ 
  2454138.551 & 2007 Feb  7
 & HST & WFPC2 & -0.22173  &  0.26308  &  0.00146  &  0.00230  & 
-0.03429  & -0.53079  &  0.00497  &  0.00582  \\ 
  2454138.567 & 2007 Feb  7
 & HST & WFPC2 & -0.21978  &  0.26791  &  0.00202  &  0.00226  & 
-0.03576  & -0.52712  &  0.00270  &  0.00479  \\ 
  2454469.653 & 2008 Jan  4
 & HST & WFPC2  &  0.23786  & -1.27383  &  0.00404  &  0.00824  & 
-0.02399  & -0.28555  &  0.00670  &  0.00831  \\ 
  2454552.897 & 2008 Mar 27
 & Keck & NIRC2 &  0.19974  & -0.10941  &  0.00930  &  0.00956  & 
\nodata & \nodata & \nodata & \nodata \\
  2454556.929 & 2008 Mar 31
 & Keck & NIRC2 &  0.32988  & -0.77111  &  0.00455  &  0.00557  & 
 0.00439  & -0.76848  &  0.01239  &  0.01280  \\ 
  2454556.948 & 2008 Mar 31
 & Keck & NIRC2 &  0.33367  & -0.77427  &  0.00890  &  0.00753  & 
 0.01363  & -0.76500  &  0.01976  &  0.01252  \\ 
  2454556.964 & 2008 Mar 31
 & Keck & NIRC2 &  0.33267  & -0.77874  &  0.00676  &  0.00485  & 
 0.00576  & -0.77375  &  0.01212  &  0.01283  \\ 
  2454557.004 & 2008 Mar 31
 & Keck & NIRC2 &  0.33543  & -0.78372  &  0.00404  &  0.00592  & 
 0.00854  & -0.77313  &  0.01199  &  0.00897  \\ 
  2454557.020 & 2008 Mar 31
 & Keck & NIRC2 &  0.33491  & -0.78368  &  0.00374  &  0.00473  & 
 0.00075  & -0.76974  &  0.00907  &  0.01015  \\ 
  2454557.039 & 2008 Mar 31
 & Keck & NIRC2 &  0.33712  & -0.78464  &  0.00740  &  0.00936  & 
 0.00988  & -0.77084  &  0.01793  &  0.01543  \\ 
  2454557.058 & 2008 Mar 31
 & Keck & NIRC2 &  0.33549  & -0.78692  &  0.00868  &  0.00852  & 
 0.01533  & -0.76117  &  0.00765  &  0.01571  \\ 
  2454557.074 & 2008 Mar 31
 & Keck & NIRC2 &  0.33128  & -0.78867  &  0.01431  &  0.01411  & 
 0.00645  & -0.76297  &  0.01639  &  0.01390  \\ 
  2454557.091 & 2008 Mar 31
 & Keck & NIRC2 &  0.33687  & -0.79462  &  0.00803  &  0.00717  & 
 0.00708  & -0.76986  &  0.01532  &  0.00787  \\ 
  2454593.726 & 2008 May  7
 & HST & NICMOS & -0.18297  &  1.08994  &  0.00354  &  0.00425  & 
 0.00243  & -0.75878  &  0.00576  &  0.00761  \\ 
  2454600.192 & 2008 May 13
 & HST & WFPC2 &  0.10847  &  0.17074  &  0.00508  &  0.00427  & 
-0.02325  &  0.19934  &  0.00480  &  0.01161  \\ 
  2454601.990 & 2008 May 15
 & HST & WFPC2 &  0.18374  & -0.13041  &  0.00729  &  0.00504  & 
-0.02293  &  0.50217  &  0.00618  &  0.00614  \\ 
  2454603.788 & 2008 May 17
 & HST & WFPC2 &  0.24918  & -0.43962  &  0.00207  &  0.00574  & 
-0.01174  &  0.59613  &  0.00366  &  0.00485  \\ 
  2454605.788 & 2008 May 19
 & HST & WFPC2 &  0.29818  & -0.75412  &  0.00467  &  0.00966  & 
 0.00006  &  0.29915  &  0.00425  &  0.00613  \\ 

\enddata
 
\tablecomments{Summary of observations of the astrometric positions of Hi'iaka (H) and Namaka (N) relative to Haumea. The difference in brightness ($\sim$6) and orbital planes allow for 
a unique identification of each satellite without possibility of confusion. The method for obtaining the astrometric positions and errors is described in Section
\ref{observationsection} and \citet{2005ApJ...632L..45B}. On a few dates, the fainter Namaka was not detected because the observations were not of sufficiently deep or Namaka was located within the PSF of Haumea. This data is shown graphically in Figure \ref{modelobs} and the residuals to the fit shown in Figure \ref{residuals}. For reasons described in the text, only the HST data is used to calculate the orbital parameters, which are shown in Table \ref{orbparams}.
 }

\end{deluxetable*}


\label{observationsection}
Observations from Keck are reduced as in B05. Known bad pixels were interpolated over and each image divided by a median flat-field. The images were then pair-wise 
subtracted (from images taken with the same filter). The astrometric centroid of each of the visible objects is determined by fitting two-dimensional Gaussians. Converting image distance to on-the-sky astrometric distance is achieved using the recently derived pixel scale of \citet{2008ApJ...689.1044G}, who calibrate the absolute astrometry of the NIRC2 camera and find a plate scale of 0.009963"/pixel (compared to the previously assumed value of 0.009942"/pixel) and an additional rotation of 0.13$^{\circ}$ compared with the rotation information provided in image headers. \citet{2008ApJ...689.1044G} and \citet{2008arXiv0807.4139H} find that the plate-scale and rotation are stable over the timescale of our observations. Error bars are determined from the scatter of the measured distances from each individual image; typical integration times were about 1 minute. When the inner satellite is not detected in individual images, but can be seen in the stacked image, then the position is taken from the stacked image, after individually rotating, and the error bars 
are simply scaled to the error bars of the outer satellite by multiplying by the square root of the ratio of 
signal/noise ($\sim$5). The minute warping of the NIRC2 
fields\footnote{See the NIRC2 Astrometry page at \texttt{http:// www2.keck.hawaii.edu/inst/nirc2/forReDoc/ post\_observing/dewarp/}.} is much smaller than the quoted error bars. 

HST benefits from a known and stable PSF and well-calibrated relative astrometry. This allows for precise measurements, even when the satellites are quite close to Haumea. For each of the HST observations, model PSFs were generated using Tiny Tim\footnote{\texttt{http://www.stsci.edu/software/tinytim/tinytim.html}.}. The model PSFs assumed solar colors, as appropriate for Haumea and its satellites, and were otherwise processed according to the details given in \emph{The Tiny Tim User's Guide}. All three PSFs were then fitted simultaneously to minimize $\chi^2$, with errors taken from photon and sky noise added in quadrature. Bad pixels and cosmic rays were identified by hand and masked out of the $\chi^2$ determination. The distortion correction of \citet{2003PASP..115..113A} for WFPC2 is smaller than our error bars for our narrow angle astrometry and was not included. Relative on-the-sky positions were calculated using the \texttt{xyad} routine of the IDL Astro Library, which utilizes astrometry information from the image headers.

The acquisition and analysis of the satellite images taken in 2005 at Keck are described in B05. However, there is 
a sign error in the R. A. Offsets listed in Table 1 of 2005; the values listed are actually the on-the-sky deviations 
(as visible from their Figure 1). Despite this typographical error, the fit of B05 was carried out correctly. The observed locations 
and estimated errors of the inner satellite are given in \citet{2006ApJ...639L..43B}. The astrometric positions reported in Table \ref{obs} are slightly different based on 
a reanalysis of some of the data as well as a new plate scale and rotation, discussed above. Based on our orbital solution and a reinvestigation of the 
images, we have determined that the May 28, 2005 observation of Namaka reported in \citet{2006ApJ...639L..43B} was spurious; residual long-lived speckles from the adaptive optics 
correction are often difficult to distinguish from faint close-in satellites.

In 2006, HST observed Haumea with the High Resolution Camera of the Advanced 
Camera for Surveys (Program 10545). Two five minute integrations were taken at the beginning and end of a single orbit. The raw images were
used for fitting, requiring distorted PSFs and distortion-corrected astrometry. The astrometric accuracy of ACS is estimated to be $\sim$0.1 pixels
 to which we add the photon noise error in the positions of the three 
objects. The high precision of ACS allows for motion to be detected between these two exposures, so these errors are not based on 
the scatter of multiple measurements as with all the other measurements.

At the beginning of February 2007, Hubble observed Haumea for 5 orbits, obtaining highly accurate positions for both 
satellites (Program 10860). The motion of the satellites from orbit to orbit is easily detected, and motion during a single orbit can even be 
significant, so we subdivided these images into 10 
separate ``observations''. The timing of the observations were chosen to have a star in the field of view, from 
which the Tiny Tim PSF parameters are modeled in manner described in \citet{2004AJ....127.2413B}. The observations do not track Haumea, but are fixed on the star to get the best PSF which is 
then appropriately smeared for the motion of the objects. Even though these observations were taken with the Wide Field 
Planetary Camera --- the ACS High-Resolution Camera failed only a week earlier --- the PSF fitting works excellently and 
provides precise positions. Astrometric errors for these observations were determined from the observed scatter in positions after subtracting
the best fit quadratic trend to the data, so that observed orbital motion is not included in the error estimate. We note here that 
combined deep stacks of these images revealed no additional outer satellites brighter than $\sim$0.25\% fractional brightness at distances out to about a tenth of the Hill sphere
(i.e. about 0.1\% of the volume where additional satellites would be stable). 

In 2008, we observed Haumea with Keck NIRC2 on the nights of March 28 and March 31. The observations on March 31 in H band lasted for
about 5 hours under good conditions, with clear detections of both satellites in each image. These were processed as described above. 
Observations where Haumea had a large FWHM were removed; about 75\% of the data was kept.
As with the February 2007 HST data, we divided the observations into 10 separate epochs and determined errors from scatter after subtracting a quadratic trend. The motion of the outer satellite
is easily detected, but the inner satellite does not move (relative to Haumea) within the errors because it is at southern elongation. The March 28 data was not nearly as good as the March 31 data due to poor weather conditions and only the outer satellite is clearly detected. 

In early May 2008, HST observed Haumea using the NICMOS camera (Program 11169). These observations were processed as described above, though a few images with 
obvious astrometric errors (due to the cosmic rays which riddle these images) were discarded. These are the same observations discussed by \citet{fraserbrowninprep}.

In mid-May 2008, we observed Haumea at five epochs using the Wide Field Planetary Camera (WFPC2), over the course of 8 days (Program 11518). Each of 
these visits consisted of four $\sim$10 minute exposures. These data, along with an observation in January 2008, were processed as described above. Although 
we expect that some of these cases may have marginally detected motion of the satellites between the four exposures, ignoring the motion only has the 
effect of slightly inflating the error bars for these observations. Namaka was too close to Haumea ($\lesssim 0.1$") to observe in the May 12, 2008 
image, which is not used.

The derived on-the-sky relative astrometry for each satellite, along with the average Julian Date of the observation and other information are summarized in Table \ref{obs}. These are the astrometric data used for orbit fitting in this paper. In earlier attempts to determine the orbit of Namaka, we also obtained other observations. On the nights of April 20 and 21, 2006, we observed Haumea with the OSIRIS camera and LGSAO at Keck. Although OSIRIS is an integral-field spectrometer, our observations were taken in photometric mode. In co-added images, both satellites were detected on both nights. We also received queue-scheduled observations of Haumea with the NIRI camera on Gemini and the LGSAO system Altair. In 2007, our Gemini program resulted in four good nights of data on April 9 and 13, May 4, and June 5. In 2008, good observations were taken on April 20, May 27, and May 28. In each of the Gemini images, the brighter satellite is readily found, but the fainter satellite is often undetectable. 

The accuracy of the plate scale and rotation required for including OSIRIS and Gemini observations is unknown, so these data are not used for orbit determination. We have, however, projected the orbits derived below to the positions of all known observations. The scatter in the Monte Carlo orbital suites (described below) at the times of these observations is small compared to the astrometric error bars of each observation, implying that these observations are not important for improving the fit. Predicted locations do not differ significantly from the observed locations, for any observation of which we are aware, including those reported in \citet{2006ApJ...640L..87B} and \citet{2008arXiv0811.3732L}.

\label{satphoto}
Using these observations, we can also do basic relative photometry of the satellites. The brightness of the satellites was computed from the height of the best-fit PSFs found to match the May 15, 2008 HST WFPC2 observation. Based on the well-known period and phase of the light curve of Haumea (\citealp{2008AJ....135.1749L}; D. Fabrycky, pers. comm.), Haumea was at its faintest during these observations and doesn't change significantly in brightness. Hi'iaka was found to be $\sim$10 times fainter than Haumea and Namaka $\sim$3.7 times fainter than Hi'iaka. 

\section{Orbit Fitting and Results}
\label{orbitfitting}
The orbit of Hi'iaka and mass of Haumea were originally determined by B05. From three detections of Namaka, \citet{2006ApJ...639L..43B} estimated three possible orbital periods around 18, 19, and 35 days. The ambiguity resulted from an under-constrained problem: at least 4-5 astrometric observations are required to fully constrain a Keplerian orbit. Even after additional astrometry was obtained, however, no Keplerian orbit resulted in a reasonable fit, where, as usual, goodness-of-fit is measured by the $\chi^2$ statistic, and a reduced $\chi^2$ of order unity is required to accept the orbit model. By forward integration of potential Namaka orbits, we confirmed that non-Keplerian perturbations due to Hi'iaka (assuming any reasonable mass) causes observationally significant deviations in the position of Namaka on timescales much longer than a month. Therefore, we expanded our orbital model to include fully self-consistent three-body perturbations. 

\subsection{Three Point-Mass Model}

Determining the orbits and masses of the full system requires a 15-dimensional, highly non-linear, global $\chi^2$ minimization. We found this to be impractical without a good initial guess for the orbit of Namaka to reduce the otherwise enormous parameter space, motivating the acquisition of multiple observations within a short enough timescale that Namaka's orbit is essentially Keplerian. Fitting the May 2008 HST data with a Keplerian model produced the initial guess necessary for the global minimization of the fully-interacting three point-mass model. The three point-mass model uses 15 parameters: the masses of the Haumea, Hi'iaka, and Namaka, and, for both orbits, the osculating semi-major axis, eccentricity, inclination, longitude of the ascending node, argument of periapse, and mean anomaly at epoch HJD 2454615.0 (= May 28.5, 2008). All angles are defined in the J2000 ecliptic coordinate system. Using these orbital elements, we constructed the Cartesian locations and velocities at this epoch as an initial condition for the three-body integration. Using a sufficiently small timestep ($\sim$300 seconds for the final iteration), a FORTRAN 90 program integrates the system to calculate the positions relative to the primary and the positions at the exact times of observation are determined by interpolation. (Observation times were converted to Heliocentric Julian Dates, the date in the reference frame of the Sun, to account for light-travel time effects due to the motion of the Earth and Haumea, although ignoring this conversion does not have a significant effect on the solution.) Using the JPL HORIZONS ephemeris for the geocentric position of Haumea, we vectorially add the primary-centered positions of the satellites and calculate the relative astrometric on-the-sky positions of both satellites. This model orbit is then compared to the data by computing $\chi^2$ in the normal fashion. We note here that this model does not include gravitational perturbations from the Sun or center-of-light/center-of-mass corrections, which are discussed below.

Like many multi-dimensional non-linear minimization problems, searching for the best-fitting parameters required a global minimization algorithm to escape the ubiquitous local minima. Our algorithm for finding the global minimum starts with thousands of local minimizations, executed with the Levenberg-Marquardt algorithm \texttt{mpfit}\footnote{An IDL routine available at \texttt{http:// www.physics.wisc.edu/$\sim$craigm/idl/fitting.html}.} using numerically-determined derivatives. These local minimizations are given initial guesses that cover a very wide range of parameter space. Combining all the results of these local fits, the resultant parameter vs. $\chi^2$ plots showed the expected parabolic shape (on scales comparable to the error bars) and these were extrapolated to the their minima. This process was iterated until a global minimum is found; at every step, random deviations of the parameters were added to the best-fit solutions, to ensure a full exploration of parameter space. Because many parameters are highly correlated, the ability to find the best solutions was increased significantly by adding correlated random deviations to the fit parameters as determined from the covariance matrix of the best known solutions. We also found it necessary to optimize the speed of the evaluating of $\chi^2$ from the 15 system parameters; on a typical fast processor this would take a few hundredths of a second and a full local minimization would take several seconds. 

To determine the error bars on the fit parameters, we use a Monte Carlo technique (B05), as suggested in \citet{1992nrfa.book.....P}. Synthetic data sets are constructed by adding independent Gaussian errors to the observed data. The synthetic data sets are then fit using our global minimization routine, resulting in 86 Monte Carlo realizations; four of the synthetic data-sets did not reach global minima and were discarded, having no significant affect on the error estimates. One-sigma one-dimensional error bars for each parameter are given by the standard deviation of global-best parameter fits from these synthetic datasets. For each parameter individually, the distributions were nearly Gaussian and were centered very nearly on the best-fit parameters determined from the actual data. The error bars were comparable to error bars estimated by calculating where $\chi^2$ increased by 1 from the global minimum \citep[see][]{1992nrfa.book.....P}.

First, we consider a solution using only the observations from HST. Even though these are taken with different instruments (ACS, NICMOS, and mostly WFPC2), the extensive calibration of these cameras allows the direct combination of astrometry into a single dataset. The best-fit parameters and errors are shown in Table \ref{orbparams}. The reduced chi-square for this model is $\chi^2_{red}=0.64$ ($\chi^2=36.4$ with 57 degrees of freedom). The data are very well-fit by the three point mass model, as shown in Figures \ref{modelobs} and \ref{residuals}. A reduced $\chi^2$ less than 1 is an indication that error bars are overestimated, assuming that they are independent; we note that using 10 separate ``observations'' for the Feb 2007 data implies that our observations are not completely independent. Even so, $\chi^2_{red}$ values lower than 1 are typical for this kind of astrometric orbit fitting \citep[e.g.,][]{2008arXiv0812.3126G}.
Each of the fit parameters is recovered, though the mass of Namaka is only detected with a 1.2-$\sigma$ significance. Namaka's mass is the hardest parameter to determine since it requires detecting minute non-Keplerian perturbations to orbit of the more massive Hi'iaka. The implications of the orbital state of the Haumea system are described in the next section. We also list in Table \ref{state} the initial condition of the three-body integration for this solution.

\begin{figure}
\begin{center}
\includegraphics[angle=90,width=3in,trim=0in 0in 0in -1.5in]{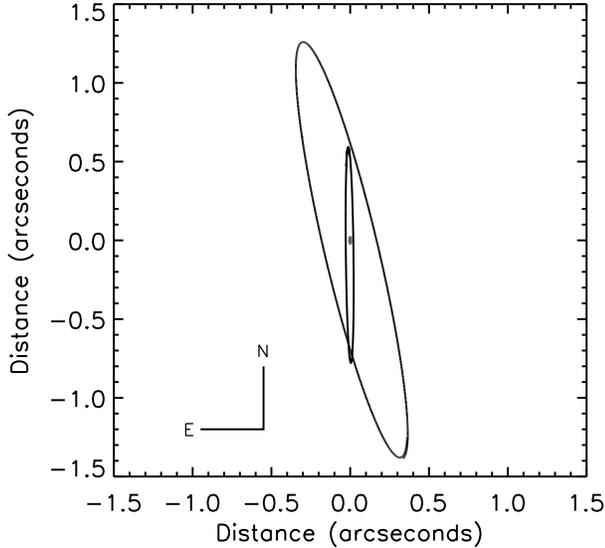}
   \caption{\label{onthesky} Relative positions of the satellites as 
viewed from Earth. The outer orbit corresponds to the brighter Hi'iaka and 
the inner orbit corresponds to the fainter Namaka. In the 
center is Haumea, drawn to scale, assuming an ellipsoid cross-section of  
500 x 1000 km \citep{2006ApJ...639.1238R} with the long axis oriented North-South. The apparent orbit changes due to 
parallax and three-body effects; this is the view near March 2008. See Figure 
\ref{modelobs} for model and data positions throughout the observation period 
(2005-2008).}
\end{center}
\end{figure}

\begin{figure*}
\begin{center}
\includegraphics[angle=90,width=6in,trim=0in 0in 0in -1.5in]{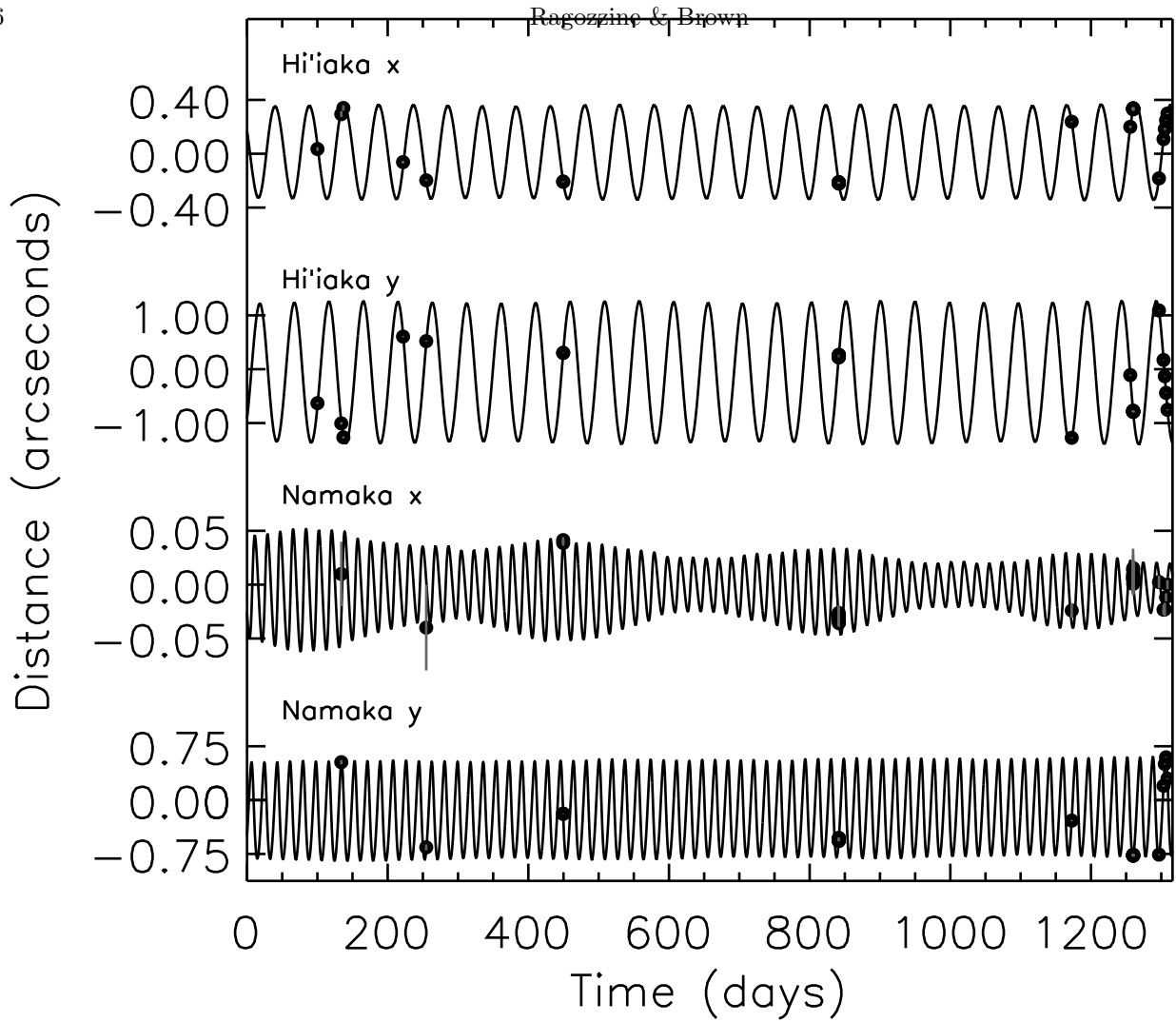}
   \caption{\label{modelobs} Observed positions and model positions of Hi'iaka and 
Namaka. From top to bottom, the curves represent the model on-the-sky position of Hi'iaka 
in the $x$-direction (i.e. the negative offset in Right Ascension), Hi'iaka in the 
$y$-direction (i.e. the offset in Declination), Namaka in the $x$-direction, and Namaka in 
the $y$-direction, all in arcseconds. Points represent astrometric observations as reported in Table 
\ref{obs}. Error bars are also shown as gray lines, but are usually much smaller than the 
points. The three-point mass model shown here is fit to the HST-data only, with a reduced 
$\chi^2_{\rm red}$ of 0.64. The residuals for this solution are shown in Figure \ref{residuals}.
Note that each curve has its own scale bar and that the curves are
offset for clarity. The model is shown for 1260 days, starting on HJD 2453297.0, $\sim$100 days 
before the first observation and ending just after the last observation. Visible are the orbital 
variations ($\sim$49.5 days for 
Hi'iaka and $\sim$18.6 days for Namaka), the annual variations due to Earth's parallax, 
and an overall trend due to a combination of Haumea's orbital motion and the precession of Namaka's 
orbit.
}
\end{center}
\end{figure*}

The HST data are sufficient to obtain a solution for Hi'iaka's orbit that is essentially the same as the orbit obtained from the initial Keck data in B05. Nevertheless, the amount and baseline of Keck NIRC2 data is useful enough to justify adding this dataset to the fit. Simply combining these datasets and searching for the global minimum results in a significant degradation in the fit, going from a reduced $\chi^2$ of 0.64 to a reduced $\chi^2$ of $\sim$1.10, although we note that this is still an adequate fit. Adding the Keck data has the effect of generally lowering the error bars and subtly changing some of the retrieved parameters. Almost all of these changes are within the $\sim$1-$\sigma$ error bars of the HST only solution, except for the mass estimate of Namaka. Adding the Keck data results in a best-fit Namaka mass a factor of 10 lower than the HST data alone. The largest mass retrieved from the entire Monte Carlo suite of solutions to the HST+Keck dataset is $\sim8 \x \e{17}$ kg, i.e. a Namaka/Haumea mass ratio of $2 \x \e{-4}$, which is inconsistent with the brightness ratio of $\sim$0.02, for albedos less than 1 and densities greater than 0.3 g/cc. However, this solution assumes that the Keck NIRC2 absolute astrometry (based on the solution of \citet{2008ApJ...689.1044G}, which is not directly cross-calibrated with HST) is perfectly consistent with HST astrometry. In reality, a small difference in the relative plate scale and rotation between these two telescopes could introduce systematic errors. Adding fitted parameters that adjust the plate scale and rotation angle does not help, since this results in over-fitting, as verified by trial fitting of synthetic datasets. We adopt the HST-only solution, keeping in mind that the nominal mass of Namaka may be somewhat overestimated.

Using the Monte Carlo suite of HST-only solutions, we can also calculate 
derived parameters and their errors. Using Kepler's Law (and ignoring the 
other satellite), the periods of Hi'iaka and Namaka are 49.462 $\pm$ 0.083 
days and 18.2783 $\pm$ 0.0076 days, respectively, with a ratio of 2.7060 
$\pm$ 0.0037, near the 8:3 resonance. The actual mean motions (and resonance occupation)
will be affected by the presence of the other satellite and the non-spherical 
nature of the primary (discussed below). 

The mass ratios of the satellite to Haumea are 0.00451 $\pm$ 0.00030 and 
0.00051 $\pm$ 0.00036, respectively and the Namaka/Hi'iaka mass ratio is 
0.116 $\pm$ 0.086. The mutual inclination of the two orbits is $\phi = 
13.41^{\circ} \pm 0.08^{\circ}$, where the mutual inclination is the 
actual angle between the two orbits, given by $\cos \phi = \cos i_H \cos 
i_N + \sin i_H \sin i_N \cos (\Omega_H - \Omega_N)$, where $i$ and 
$\Omega$ are the inclination and longitude of ascending node. The origin 
of this significantly non-zero mutual inclination is discussed in Section 
\ref{tidalinc}. The mean longitude, $\lambda \equiv \Omega + \omega + M$,
is the angle between the reference line (J2000 ecliptic first point of Ares)
and is determined well; the errors in the argument of periapse ($\omega$) and 
mean anomaly ($M$) shown in Tabel \ref{orbparams} are highly anti-correlated. 
Our Monte Carlo results give $\lambda_H = 153.80 \pm 0.34$ degrees and
$\lambda_N = 202.57 \pm 0.73$ degrees. Finally, under the nominal point-mass model, Namaka's argument of periapse
changes by about -6.5$^{\circ}$ per year during the course of the observations,
implying a precession period of about 55 years; the non-Keplerian nature of Namaka's 
orbit is detected with very high confidence.


\begin{deluxetable*}{llrcrl}
\tablecaption{Fitted Parameters of the Haumea System \label{orbparams}}
\tablehead{
\colhead{Object} & \colhead{Parameter} & \colhead{Value} & \colhead{} & \colhead{Error} & \colhead{Units}  }

\startdata

Haumea  & Mass & 4.006 & $\pm$ & 0.040 & $10^{21}$ kg \\
Hi'iaka & Mass & 1.79 & $\pm$ & 0.11 & $10^{19}$ kg \\ 
        & Semi-major axis & 49880 & $\pm$ & 198 & km \\
        & Eccentricity & 0.0513 & $\pm$ & 0.0078 & \\
        & Inclination & 126.356 & $\pm$ & 0.064 & degrees \\
        & Longitude of ascending node &  206.766 & $\pm$ & 0.033 & degrees \\   
        & Argument of periapse & 154.1 & $\pm$ & 5.8 & degrees \\
        & Mean anomaly & 152.8 & $\pm$ & 6.1 & degrees \\
Namaka  & Mass & 1.79 & $\pm$ & 1.48 & $10^{18}$ kg \\ 
        & Semi-major axis & 25657 & $\pm$ & 91 & km \\
        & Eccentricity & 0.249 & $\pm$ & 0.015 & \\
        & Inclination & 113.013 & $\pm$ & 0.075 & degrees \\
        & Longitude of ascending node & 205.016 & $\pm$ & 0.228 & degrees \\   
        & Argument of periapse & 178.9 & $\pm$ & 2.3 & degrees \\
        & Mean anomaly & 178.5 & $\pm$ & 1.7 & degrees \\

\enddata
 
\tablecomments{Orbital parameters at epoch HJD 2454615.0. The nominal values are from the best fit to the HST data, while the (often correlated) error bars are the standard deviation of fitted values returned from a Monte Carlo suite of 86 datasets. These are the osculating orbital elements at this epoch; due to the three-body interactions, these values (especially the angles) change over the timescale of observations. All angles are referenced to the J2000 ecliptic coordinate system. See Table \ref{state} for the Cartesian positions of the two satellites at this epoch.}

\end{deluxetable*}

\begin{deluxetable*}{lrrrrrr}
\tablecaption{State Vector for the Haumea System \label{state}}
\tablehead{
\colhead{Object} & \colhead{$x$ (m)} & \colhead{$y$ (m)} & \colhead{$z$ (m)} & \colhead{$v_x$ (m/s)} & \colhead{$v_y$ (m/s)} & \colhead{$v_z$ (m/s)}  }

\startdata
Hi'iaka & -18879430 &       -36260639  &       -32433454  &              60.57621  &               1.85403  &             -34.81242  \\
Namaka & -28830795  &       -13957217  &        -1073907  &              16.07022  &             -26.60831  &             -72.76764  \\ 
\enddata
 
\tablecomments{Cartesian position and velocity of Haumea's satellites relative to Haumea in the J2000 ecliptic coordinate system at epoch HJD 2454615.0 corresponding to the best-fit orbital parameters shown in Table \ref{orbparams}.}

\end{deluxetable*}


\subsection{Including the $J_2$ of Haumea}

The non-spherical nature of Haumea can introduce additional, potentially observable, non-Keplerian effects. The largest of these effects is due to the lowest-order gravitational moment, the quadrupole term (the dipole moment is 0 in the center of mass frame), described by $J_2$ \citep[see, e.g.,][]{2000ssd..book.....M}. Haumea rotates over 100 times during a single orbit of Namaka, which orbits quite far away at $\sim$35 primary radii. To lowest order, therefore, it is appropriate to treat Haumea as having an ``effective'' time-averaged $J_2$. Using a code provided by E. Fahnestock, we integrated trajectories similar to Namaka's orbit around a homogeneous rotating tri-axial ellipsoid and have confirmed that the effective $J_2$ model deviates from the full model by less than half a milliarcsecond over three years. 

The value of the effective $J_2$ ($\equiv -C_{20}$) for a rotating homogeneous tri-axial ellipsoid was derived by \citet{1994Icar..110..225S}:
\be
J_2 R^2 = \frac{1}{10}(\alpha^2+\beta^2-2\gamma^2) \simeq 1.04 \x \e{11} \textrm{m}^2
\ee
where $\alpha,\beta$, and $\gamma$ are the tri-axial radii and the numerical value corresponds to a (498 x 759 x 980) km ellipsoid as inferred from photometry \citep{2006ApJ...639.1238R}. We note that the physical quantity actually used to determine the orbital evolution is $J_2 R^2$; in a highly triaxial body like Haumea, it is not clear how to define $R$, so using $J_2 R^2$ reduces confusion. If $R$ is taken to be the volumetric effective radius, then $R \simeq 652$ km and the $J_2 \simeq 0.244$. Note that the calculation and use of $J_2$ implicitly requires a definition of the rotation axis, presumed to be aligned with the shortest axis of the ellipsoid. 

Preliminary investigations showed that using this value of $J_2 R^2$ implied a non-Keplerian effect on Namaka's orbit that was smaller, but similar to, the effect of the outer satellite. The primary observable effect of both $J_2$ and Hi'iaka is the precession of apses and nodes of Namaka's eccentric and inclined orbit \citep{2000ssd..book.....M}. When adding the three relevant parameters --- $J_2 R^2$ and the direction of the rotational axis\footnote{When adding $J_2$, our three-body integration was carried out in the frame of the primary spin axis and then converted back to ecliptic coordinates.} --- to our fitting procedure, we found a direct anti-correlation between $J_2 R^2$ and the mass of Hi'iaka, indicating that these two parameters are degenerate in the current set of observations. The value of the reduced $\chi^2$ was lowered significantly by the addition of these parameters, with the F-test returning high statistical significance. However, the best fits placed the satellites on polar orbits. We have verified that the fitted values of $J_2 R^2$ and the spin pole perpendicular to the orbital poles is due to over-fitting of the data. We generated simulated observations with the expected value of $J_2 R^2$ and with the satellites in nearly equatorial orbits. Allowing the global fitter to vary all the parameters resulted in an over-fitted solution that placed the satellites on polar orbits. Hence, allowing $J_2$ and the spin pole to vary in the fit is not justified; the effect of these parameters on the solution are too small and/or too degenerate to detect reliably. Furthermore, since the model without these parameters already had a reduced $\chi^2$ less than 1, these additional parameters were not warranted in the first place.

It is interesting, however, to consider how including this effect would change the determination of the other parameters. We therefore ran an additional set of models with a fixed $J_2 R^2$ and fixing the spin pole (more accurately, the axis by which $J_2$ is defined) as the mass-weighted orbital pole of Hi'iaka and Namaka; since Hi'iaka is $\sim$10 times more massive than Namaka, this puts Hi'iaka on a nearly equatorial orbit ($i \simeq 1^{\circ}$), as would be expected from collisional formation. Holding $J_2 R^2$ fixed at $1.04 \x \e{11}$ m$^2$, we reanalyzed the HST dataset using our global fitting routine. As expected, none of the parameters change by more than 1-$\sigma$, except for the mass of Hi'iaka, which was reduced by almost 30\% to $\sim1.35 \x \e{19}$ kg. (In fits where $J_2 R^2$ was allowed to vary, the tradeoff between $J_2 R^2$ and Hi'iaka's mass was roughly linear, as would be expected if the sum of these effects were forced to match the observationally determined precession of Namaka.) The data were well fit by the non-point mass model, with the forced $J_2 R^2$ and spin pole solution reaching a global reduced $\chi^2$ of 0.72. Since Haumea's high-amplitude light-curve indicates a primary with a large quadrupole component, the nominal mass of Hi'iaka in the point-mass case is almost certainly an overestimate of its true mass. More data will be required to disentangle the degeneracy between the mass of Hi'iaka and the $J_2$ of Haumea. Including $J_2$ and/or the Keck data do not improve the estimates of Namaka's mass.

\begin{figure}
\begin{center}
\includegraphics[angle=90,width=3in, trim=0in 0in 0in 0.5in]{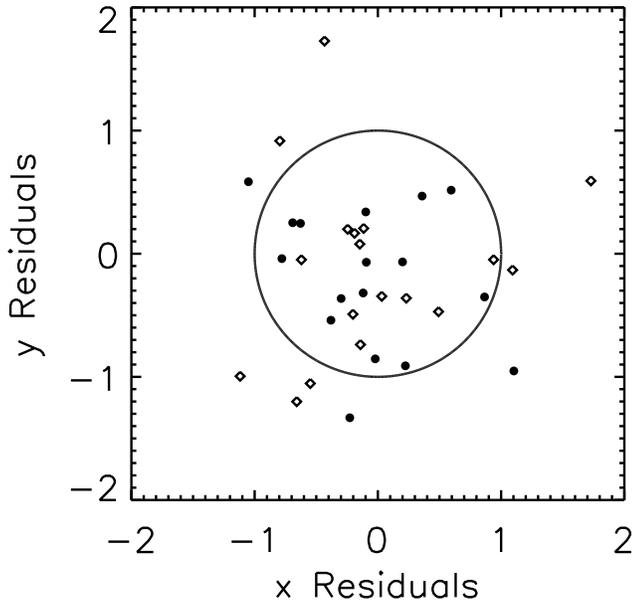}
   \caption{\label{residuals} Normalized residuals of the three-point mass fit to HST-data only. Plotted is
$(\Delta x_{\rm mod}-\Delta x_{\rm obs})/\sigma_{\Delta x}$ versus $(\Delta y_{\rm mod}-\Delta y_{\rm obs})/\sigma_{\Delta y}$
for Hi'iaka (diamonds) and Namaka (circles). Points that lie within the circle indicate where the model and observations 
vary by less than 1 error bar. See also Figure \ref{modelobs}. The residuals are roughly evenly spaced and favor neither Hi'iaka nor Namaka, implying that 
there are no major systematic effects plaguing the three-body fit. As reported in the text this solution has a reduced $\chi^2_{\rm red}$ of 0.64. 
}
\end{center}
\end{figure}

\section{Implications of Orbital Solutions}

Taking the orbital solutions derived in the previous section, we can begin to answer questions relevant to the formation and evolution of this unique satellite system.

\subsection{Mutual Events and Satellite Sizes}

According to the orbit solution, the Haumea system is currently undergoing mutual events, as reported in \citep{2008IAUC.8949....1F}. (This is also true using the other orbit solutions, e.g. HST+Keck, with or without $J_2$.) Using the known orbit, the angle between Namaka, 
Haumea, and the Earth (in the case of occultations) or the Sun (in the case of shadowing) falls well below the $\sim$13 milliarcseconds ($\sim$500 km) of the projected shortest axis of Haumea. Observing multiple mutual events can yield accurate and useful measurements of several system properties as shown by the results of the Pluto-Charon mutual event season \citep[e.g.][]{1997plch.book...85B}. The depth of an event where Namaka occults Haumea leads to the ratio of albedos and, potentially, a surface albedo map of Haumea, which is known to exhibit color variations as a function of rotational phase, indicative of a variegated surface \citep{2008AJ....135.1749L,2008arXiv0811.3732L}. Over the course of a single season, Namaka will traverse several chords across Haumea allowing for a highly accurate measurement of Haumea's size, shape, and spin pole direction \citep[e.g.,][]{2008Icar..196..578D}. The precise timing of mutual events will also serve as extremely accurate astrometry, allowing for an orbital solution much more precise than reported here. We believe that incorporating these events into our astrometric model will be sufficient to independently determine the masses of all three bodies and $J_2 R^2$. Our solution also predicts a satellite-satellite mutual event in July 2009 --- the last such event until the next mutual event season begins around the year 2100. Our knowledge of the state of the Haumea system will improve significantly with the observation and analysis of these events. See \texttt{http://web.gps.caltech.edu/$\sim$mbrown/2003EL61/mutual} for up-to-date information on the Haumea mutual events. Note that both the mutual events and the three-body nature of the system are valuable for independently checking the astrometric analysis, e.g. by refining plate scales and rotations.

\begin{figure}
\begin{center}
\includegraphics[angle=90,width=3.5in,trim=0in 0in 0in -0.5in]{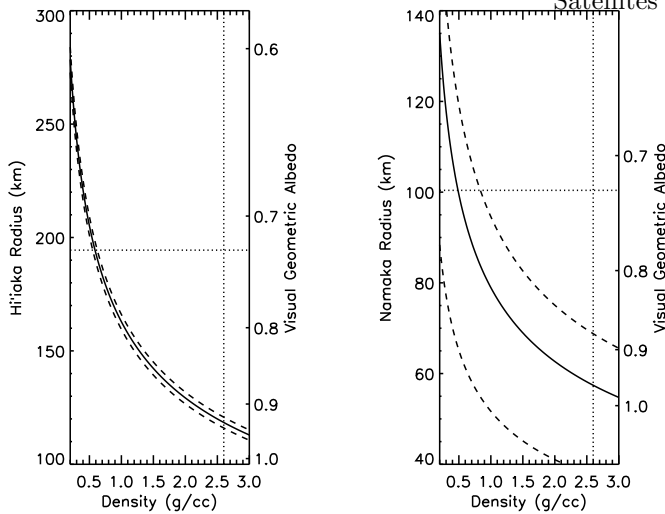}
   \caption{\label{satsalbden} Relationship between radius, density, and albedo for Hi'iaka (left) and Namaka (right). A range of possible albedos and 
densities can reproduce the determined mass and brightness ratios of Hi'iaka and Namaka, which are assumed to be spherical. The solid lines show the 
relationship for the nominal masses, reported in Table \ref{orbparams}, with dotted lines showing the 1-$\sigma$ mass error bars. Note that the mass 
of both Hi'iaka and Namaka may be overestimated (due to insufficient data, see Section \ref{orbitfitting}). The albedo and 
density of the \citet{2006ApJ...639.1238R} edge-on model for Haumea are shown by dotted lines. Both satellites must have lower 
densities and/or higher albedos than Haumea. The similar spectral \citep{2006ApJ...640L..87B} and photometric \citep{fraserbrowninprep} properties of 
Haumea, Hi'iaka, and Namaka indicate that their albedos should be similar. Under the assumption that the satellites have similar albedos to Haumea, the 
densities of the satellites indicate that they are primarily composed of water ice ($\rho \approx 1.0$ g/cc). Low satellite densities would bolster the 
hypothesis that the satellites formed from the collisional remnants of the water ice mantle of the differentiated proto-Haumea. Observation of
Haumea-Namaka mutual events will allow for much more precise and model-independent measurements of Namaka's radius, density, and albedo.} 
\end{center}
\end{figure}

Using the best-fit mass ratio and the photometry derived in Section \ref{satphoto}, we can estimate the range of albedos and densities for the two 
satellites. The results of this calculation are shown in Figure \ref{satsalbden}. The mass and brightness ratios clearly show that the satellites must 
either have higher albedos or lower densities than Haumea; the difference is probably even more significant than shown in Figure \ref{satsalbden} since 
the nominal masses of Hi'iaka and Namaka are probably overestimated (see Section \ref{orbitfitting}). The similar spectral \citep{2006ApJ...640L..87B} 
and photometric \citep{fraserbrowninprep} properties of Haumea, Hi'iaka, and Namaka indicate that their albedos should 
be similar. Similar surfaces are also expected from rough calculations of ejecta exchange discussed by \citet{2009Icar..199..571S}, though 
\citet{2008arXiv0811.2104B} provide a contrary viewpoint. If the albedos are comparable, the satellite densities indicate a 
mostly water ice composition ($\rho \approx 1.0$ g/cc). This lends support to the hypothesis that the satellites are formed from a collisional debris 
disk composed primarily of water ice from the shattered mantle of Haumea. This can be confirmed in the future with a direct measurement of Namaka's 
size from mutual event photometry. Assuming a density of water ice, the estimated radii of Hi'iaka and Namaka are $\sim$160 km and $\sim$80 km, 
respectively.

\subsection{Long-term Orbital Integrations}

It is surprising to find the orbits in an excited state, both with non-zero eccentricities and with a rather large mutual inclination. In contrast the 
regular satellite systems of the gas giants, the satellites of Mars, the three satellites of Pluto \citep{2008AJ....135..777T}, and asteroid triple 
systems with well known orbits \citep{2005Natur.436..822M} are all in nearly circular and co-planar orbits. In systems of more than one satellite, 
perturbations between the satellites produce forced eccentricities and inclinations that will remain even with significant damping. If the excited 
state of the Haumea system is just a reflection of normal interactions, then there will be small free eccentricities and inclinations, which can be 
estimated by integrating the system for much longer than the precession timescales and computing the time average of these elements. Using this 
technique, and exploring the entire Monte Carlo suite of orbital solutions, we find that the free eccentricity of Hi'iaka is $\sim$0.07, the free 
eccentricity of Namaka is $\sim0.21$, and the time-averaged mutual inclination is $\sim$12.5$^{\circ}$. Non-zero free eccentricities and inclinations imply that the excited state of the system is not due to satellite-satellite perturbations. These integrations were calculated using the n-body code SyMBA \citep{1994Icar..108...18L} using the regularized mixed variable symplectic method based on the mapping by \citet{1991AJ....102.1528W}. Integration of all the Monte Carlo orbits showed for $\sim$2000-years showed no signs of instability, though we do note that the system chaotically enters and exits the 8:3 resonance. The orbital solutions including $J_2 R^2 \simeq 1.04 \x \e{11}$ m$^2$ were generally more chaotic, but were otherwise similar to the point-mass integrations.

These integrations did not include the effect of the Sun, which adds an additional minor torque to the system that is negligible ($\Delta \Omega \sim \e{-5}$ degrees) over the timescale of observations. The effects of the sun on the satellite orbits on long-time scales were not investigated. While the relative inclination between the satellite orbits and Haumea's heliocentric orbit ($\sim$119$^{\circ}$ for Hi'iaka and $\sim$105$^{\circ}$ for Namaka) places this system in the regime where the Kozai effect can be important \citep{1962AJ.....67..591K,2008arXiv0809.2095P}, the interactions between the satellites are strong enough to suppress weak Kozai oscillations due to the Sun, which are only active in the absence of other perturbations \citep{2007ApJ...669.1298F}.

We did not include any correction to our solution for possible differences between the center of light (more precisely, the center of fitted PSF) and center of mass of Haumea. This may be important since \citet{2008AJ....135.1749L} find that Haumea's two-peaked light curve can be explained by a dark red albedo feature, which could potentially introduce an systematic astrometric error. The February 2007 Hubble data and March 2008 Keck data, both of which span a full rotation, do not require a center-of-light/center-of-mass correction for a good fit, implying that the correction should be smaller than $\sim$2 milliarcseconds (i.e. $\sim$70 km). Examination of the all the astrometric residuals and a low reduced-$\chi^2$ confirm that center-of-light/center-of-mass corrections are not significant at our level of accuracy. For Pluto and Charon, albedo features can result in spurious orbital astrometry because Pluto and Charon are spin-locked; this is not the case for Haumea. For Keck observations where Namaka is not detected (see Table \ref{obs}), it is usually because of low signal-to-noise and Namaka's calculated position is not near Haumea. In the cases where Namaka's light contaminates Haumea, the induced photocenter error would be less than the observed astrometric error.

\subsection{Tidal Evolution}

All of the available evidence points to a scenario for the formation of Haumea's satellites similar to the formation of the Earth's moon: a large oblique collision created a disk of debris composed mostly of the water ice mantle of a presumably differentiated proto-Haumea. Two relatively massive moons coalesced from the predominantly water-ice disk near the Roche lobe. Interestingly, in studying the formation of Earth's Moon, about one third of the simulations of \citet{1997Natur.389..353I} predict the formation of two moonlets with the outer moonlet $\sim$10 times more massive the inner moonlet. Although the disk accretion model used by \citet{1997Natur.389..353I} made the untenable assumption that the remnant disk would immediately coagulate into solid particles, the general idea that large disks with sufficient angular momentum could result in two separate moons seems reasonable. Such collisional satellites coagulate near the Roche lobe (e.g. a distances of $\sim$3-5 primary radii) in nearly circular orbits and co-planar with the (new) rotational axis of the primary. For Haumea, the formation of the satellites is presumably concurrent with the formation of the family billions of years ago \citep{2007AJ....134.2160R} and the satellites have undergone significant tidal evolution to reach their current orbits (B05). 

\subsubsection{Tidal Evolution of Semi-major Axes}

The equation for the typical semi-major axis tidal expansion of a single-satellite due to primary tides is \citep{2000ssd..book.....M}:
\be\label{tidala}
\dot{a}=\frac{3k_{2p}}{Q_p} q \left( \frac{R_p}{a} \right)^5 na
\ee
where $k_{2p}$ is the second-degree Love number of the primary, $Q_p$ is the primary tidal dissipation parameter \citep[see, e.g.,][]{1966Icar....5..375G}, $q$ is the mass ratio, $R_p$ is the primary radius, $a$ is the satellite semi-major axis, and $n\equiv\sqrt{\frac{GM_{tot}}{a^3}}$ is the satellite mean motion. As pointed out by B05, applying this equation to Hi'iaka's orbit (using the new-found $q = 0.0045$) indicates that Haumea must be extremely dissipative: $Q_p \simeq 17$, averaged over the age of the solar system, more dissipative than any known object except ocean tides on the present-day Earth. This high dissipation assumes an unrealistically high $k_2 \simeq 1.5$, which would be achieved only if Haumea were perfectly fluid. Using the strength of an rocky body and the \citet{1995geph.conf....1Y} method of estimating $k_2$, Haumea's estimated $k_{2p}$ is $\sim0.003$. Such a value of $k_{2p}$ would imply a absurdly low and physically implausible $Q_p \ll 1$. Starting Hi'iaka on more distant orbits, e.g. the current orbit of Namaka, does not help much in this regard since the tidal expansion at large semi-major axes is the slowest part of the tidal evolution. There are three considerations, however, that may mitigate the apparent requirement of an astonishingly dissipative Haumea. First, if tidal forcing creates a tidal bulge that lags by a constant time \citep[as in][]{MignardII}, then Haumea's rapid rotation (which hardly changes throughout tidal evolution) would naturally lead to a significant increase in tidal evolution. In other words, if $Q_p$ is frequency dependent \citep[as it seems to be for solid bodies, see][]{2007JGRE..11212003E}), then an effective $Q$ of $\sim$16 may be equivalent to an object with a one-day rotation period maintaining an effective $Q$ of 100, the typically assumed value for icy solid bodies. That is, Haumea's higher-than-expected dissipation may be related to its fast rotation. Second, the above calculation used the volumetric radius $R_p \simeq 650$ km, in calculating the magnitude of the tidal bulge torque, where we note that Equation \ref{tidala} assumes a spherical primary. A complete calculation of the actual torque caused by tidal bulges on a highly non-spherical body is beyond the scope of this paper, but it seems reasonable that since tidal bulges are highly distance-dependent, the volumetric radius may lead to an underestimate in the tidal torque and resulting orbital expansion. Using $R_p \simeq 1000$ km, likely an overestimate, allows $k_{2p}/Q_p$ to go down to $7.5 \x \e{-6}$, consistent both with $k_{2p} \simeq 0.003$ and $Q_p \simeq 400$. Clearly, a reevaluation of tidal torques and the resulting orbital change for satellites around non-spherical primaries is warranted before making assumptions about the tidal properties of Haumea. 

The third issue that affects Hi'iaka's tidal evolution is Namaka. While generally the tidal evolution of the two satellites is independent, if the satellites form a resonance it might be possible to boost the orbital expansion of Hi'iaka via angular momentum transfer with the more tidally affected Namaka. (Note that applying the semi-major axis evolution questions to Namaka requires a somewhat less dissipative primary, i.e. $Q_p$ values $\sim$8 times larger than discussed above.) Even outside of resonance, forced eccentricities can lead to higher dissipation in both satellites, somewhat increasing the orbital expansion rate. Ignoring satellite interactions and applying Equation \ref{tidala} to each satellite results in the expected relationship between the mass ratio and the semi-major axis ratio of \citep{1999AJ....117..603C,2000ssd..book.....M}:
\be\label{masssma}
\frac{m_1}{m_2} \simeq \left( \frac{a_1}{a_2} \right)^{13/2}
\ee
where evaluating the right-hand side using the determined orbits implies that $\frac{m_1}{m_2} \simeq 75.4 \pm 0.4$. This mass ratio is highly inconsistent with a brightness ratio of $\sim$3.7 for the satellites, implying that the satellites have not reached the asymptotic tidal end-state. Equation \ref{masssma} is also diagnostic of whether the tidally-evolving satellites are on converging or diverging orbits: that the left-hand side of Equation \ref{masssma} is greater than the right-hand side implies that the satellites are on convergent orbits, i.e. the ratio of semi-major axes is  increasing and the ratio of the orbital periods is decreasing (when not in resonance).

\subsubsection{Tidal Evolution of Eccentricities and Inclinations}
Turning now from the semi-major axis evolution, we consider the unexpectedly large eccentricities and non-zero inclination of the Namaka and Hi'iaka. None of the aforementioned considerations can explain the highly excited state of the Haumea satellite system. As pointed out by B05, a simple tidal evolution model would require that the eccentricities and, to some extent, inclinations are significantly damped when the satellites are closer to Haumea, as eccentricity damping is more efficient than semi-major axis growth. One would also expect that the satellites formed from a collision disk with low relative inclinations. So, while a mutual inclination of 13$^{\circ}$ is unlikely to occur by random capture, a successful model for the origin of the satellites must explain why the satellites are relatively far from co-planar.

The unique current orbital state of the Haumea system is almost certainly due to a unique brand of tidal evolution. Terrestrial bodies are highly dissipative, but none (except perhaps some asteroid or KBO triple systems) have large significantly interacting satellites. On the other hand, gas giant satellite systems have multiple large interacting satellites, but very low dissipation and hence slow semi-major axis change. The change in semi-major axes is important because it causes a large change in the period ratio, allowing the system to cross many resonances, which can strongly change the nature of the system and its evolution. Even though Haumea is in a distinct niche of tidal parameter space, we can gain insights from studies of other systems, such as the evolution of satellites in the Uranian system \citep[e.g.,][]{1988Icar...74..172T,1989Icar...78...63T,1988Icar...76..295D,1990Icar...85..444M} or the interactions of tidally-evolving exoplanets \citep[e.g.,][]{2002ApJ...564.1024W,2003CeMDA..87...99F}. In addition, since the results of some Moon-forming impacts resulted in the creation of two moons \citep{1997Natur.389..353I}, \citet{1999AJ....117..603C} studied the tidal evolution of an Earth-Moon-moon system, in many ways similar to the Haumea system.

Using the results of these former investigations, we can qualitatively explain the excited state of the Haumean system. As the satellites were evolving outward at different rates, the ratio of orbital periods would periodically reach a resonant ratio. For example, as the current system is in/near the 8:3 resonance, it probably passed through the powerful 3:1 resonance in the relatively recent past. Since the satellites are on convergent orbits, they would generally get caught into these past resonances, even if their early eccentricities and inclinations were low. Note that this simple picture of resonance capture must be investigated numerically \citep[as in, e.g.,][]{1988Icar...74..172T} and that higher-order resonances may act differently from lower-order resonances \citep{2008Icar..193..267Z}. Further semi-major axis growth while trapped in the resonance rapidly pumps eccentricities and/or inclinations, depending on the type of resonance \citep[e.g.,][]{2006Sci...313.1107W}. This continues until the satellites chaotically escape from the resonance. Escaping could be a result of either excitation \citep[related to secondary resonances, see][]{1989Icar...78...63T,1990Icar...85..444M} and/or chaotic instability due to overlapping sub-resonances, split by the large $J_2$ of Haumea \citep{1988Icar...76..295D}. Outside of resonances, tidal dissipation in the satellites can damp eccentricities while the satellites are close to Haumea, but at their current positions, eccentricity damping is very ineffective 
even for highly dissipative satellites. Inclination damping is generally slower than eccentricity damping and was probably small even when the satellites were much closer to Haumea. Therefore a ``recent'' excitation by passage through a resonance (possibly the 3:1) can qualitatively explain the current orbital configuration, which has not had the time to tidally damp to a more circular co-planar state. Numerical integrations will be needed to truly probe the history of this intriguing system and may be able to constrain tidal parameters of Haumea, Hi'iaka, and/or Namaka. 

Note that early in the history of this system when the satellites were orbiting at much smaller semi-major axes ($a \lesssim 10R_p$), the tri-axial nature of the primary would have been much more important and could have significantly affected the satellite orbits \citep{1994Icar..110..225S}. At some point during semi-major axis expansion secular resonances (such as the evection resonance) could also be important for exciting eccentricities and/or inclinations \citep[][]{1998AJ....115.1653T}. Finally, tides raised on Haumea work against eccentricity damping; for certain combinations of primary and satellite values of $k_2/Q$, tides on Haumea can pump eccentricity faster than it is damped by the satellites (especially if Haumea is particularly dissipative). While eccentricity-pumping tides on Haumea may help in explaining the high eccentricities, producing the large mutual inclination (which is hardly affected by any tidal torques) is more likely to occur in a resonance passage, as with Miranda in the Uranian system \citep{1989Icar...78...63T,1988Icar...76..295D}.\label{tidalinc}

\section{Conclusions}

Using new observations from the Hubble Space Telescope, we have solved for the orbits and masses of the two dynamically interacting 
satellites of the dwarf planet Haumea. A three-body model, using the parameters with errors given in Table \ref{orbparams}, provides an excellent match to the 
precise relative astrometry given in Table \ref{obs}. The orbital parameters of Hi'iaka, the outer, brighter satellite match well the orbit previously derived 
in \citet{2005ApJ...632L..45B}. The newly derived orbit of Namaka, the inner, fainter satellite has a surprisingly large eccentricity (0.249 $\pm$ 0.015) and 
mutual inclination to Hi'iaka of (13.41$^{\circ}$ $\pm$ 0.08$^{\circ}$). The eccentricities and inclinations are not due to mutual perturbations, but can be qualitatively explained
by tidal evolution of the satellites through mean-motion resonances. The precession effect of the non-spherical nature of the elongated primary, 
characterized by $J_2$, cannot be distinguished from the precession caused by the outer satellite in the current data. 

The orbital structure of Haumea's satellites is unlikely to be produced in a capture-related formation mechanism \citep[e.g.,][]{2002Natur.420..643G}. 
Only the collisional formation hypothesis (allowing for reasonable, but atypical, tidal evolution) can explain the nearly co-planar satellites 
(probably with low densities), the rapid spin and elongated shape of Haumea \citep{2006ApJ...639.1238R}, and the Haumea collisional family of icy 
objects with similar surfaces and orbits \citep{2007Nature..446..296}.

The future holds great promise for learning more about the Haumea system, as the orbital solution indicates that Namaka and Haumea are undergoing mutual 
events for the next several years. This will provide excellent observational constraints on the size, shape, spin pole, density, and internal structure 
of Haumea and direct measurements of satellite radii, densities, and albedos. There are also interesting avenues for future theoretical investigations, 
especially into the unique nature of tidal evolution in the Haumean system. These insights into the formation and evolution of the Haumean system can 
be combined with our understanding of other Kuiper belt binaries to investigate how these binaries form and to further decipher the history of the 
outer solar system.

\begin{acknowledgements}
We would like to thank Meg Schwamb, Aaron Wolf, Matt Holman, and others for valuable discussions and Eugene Fahnestock for 
kindly providing his code for gravity around tri-axial bodies. We also acknowledge observational support from Emily Schaller and Chad Trujillo. We 
especially acknowledge the encouragement and discussions of Dan Fabrycky. DR is grateful for the support of the Moore Foundation. This work was also 
supported by NASA Headquarters under the Earth and Space Sciences Fellowship and the Planetary Astronomy programs. This work is based on NASA/ESA 
Hubble Space Telescope program 11518 using additional data from programs 10545, 10860, and 11169. Support for all these programs was provided by NASA 
through grants HST-GO-10545, HST-GO-10860, HST-G0-11518, and HST-GO-11169 from the Space Telescope Science Institute (STScI), which is operated by 
the Association of Universities for Research in Astronomy, 
Inc., under NASA contract NAS 5-26555. Some data presented herein was obtained at the W. M. Keck Observatory, which is operated as a scientific 
partnership among the California Institute of Technology, the University of California, and the National Aeronautics and Space Administration. The 
observatory was made possible by the generous financial support of the W. M. Keck Foundation. The authors wish to recognize and acknowledge the very 
significant cultural role and reverence that the summit of Mauna Kea has always had within the indigenous Hawaiian community. We are most fortunate to 
have the opportunity to conduct observations of Haumea, Hi'iaka, and Namaka, which were named after Hawaiian goddesses.
\end{acknowledgements}

\bibliographystyle{../../aj}


\end{document}